\documentclass[runningheads]{llncs}
\let\llncssubparagraph\subparagraph
\let\subparagraph\paragraph
\usepackage[compact]{titlesec}
\let\subparagraph\llncssubparagraph
\usepackage{placeins}
\titleformat{\subsubsection}[hang]
  {\normalfont\normalsize\bfseries}{\thesubsubsection}{1em}{}
\usepackage{graphicx}
\usepackage{caption}
\usepackage[caption=false]{subfig}
\usepackage[utf8]{inputenc}
\usepackage{geometry}
\usepackage{array}
\usepackage{comment}
\usepackage[table]{xcolor}

\newcolumntype{s}{>{\columncolor[HTML]{AAACED}} p{1cm}}
\begin{document}
\title{An Improved EEG Acquisition Protocol Facilitates Localized Neural Activation}
%
%
\author{Jerrin Thomas Panachakel\inst{1}\orcidID{*} \and
Nandagopal Netrakanti Vinayak \inst{2}\orcidID{*} \and
Maanvi Nunna\inst{2}\orcidID{*} \and \\
A.G. Ramakrishnan\inst{1} \and
Kanishka Sharma\inst{1}}
\authorrunning{Jerrin et al.}
%
\institute{Department of Electrical Engineering, Indian Institute of Science, Bangalore, India\and
Department of Computer Science and Engineering, PES University, Bangalore, India
\\
\email{\{jerrinp,agr,kanishka\}@iisc.ac.in, \{nandagopalnv, maanvinunna\}@pesu.pes.edu}\\ $^*$Equal contribution }
\maketitle              
\begin{abstract}
This work proposes improvements in the electroencephalogram (EEG) recording protocols for motor imagery through the introduction of actual motor movement and/or somatosensory cues. The results obtained demonstrate the advantage of requiring the subjects to perform  motor actions following the trials of imagery. By introducing  motor actions in the protocol, the subjects are able to perform actual motor planning, rather than just visualizing the motor movement, thus greatly improving the ease with which the motor movements can be imagined. This study also probes the added advantage of administering somatosensory cues in the subject, as opposed to the conventional auditory/visual cues. These changes in the protocol show promise in terms of the aptness of the spatial filters obtained on the data, on application of the well-known common spatial pattern (CSP) algorithms. The regions highlighted by the spatial filters are more localized and consistent across the subjects when the protocol is augmented with somatosensory stimuli. Hence, we suggest that this may prove to be a better EEG acquisition protocol for detecting brain activation in response to intended motor commands in (clinically) paralyzed/locked-in patients.

\end{abstract}
\setcounter{secnumdepth}{3}
\section{Introduction}
Electroencephalogram (EEG) remains vastly popular for recording the activity of the brain due to its relatively low cost, high time resolution and easy availability in hospitals and research institutions. Recently, researchers have used EEG as a tool for understanding the neural response to motor commands in clinically unresponsive patients \cite{claassen2019detection}, giving impetus to the research on protocols under which EEG data is acquired. The problem we address in this work is the design of an effective protocol that improves the quality of the signal recorded from the subjects. This work proposes two approaches - use of somatosensory cues, and the enforcement of a protocol in which motor imagery of a specific limb is followed by the actual movement of the limb, in contrast to just imagining the movement, which has been used conventionally.

\section{Methods}
\subsection{EEG Recording Setup}
The EEG data was sampled at 1000 Hz from an ANT Neuro Eego\texttrademark Mylab  amplifier using the EEGCA64-500 montage and the 10/10 electrode placement system with ``CPz'' electrode as reference electrode. A 64-channel cap with electrooculogram (EOG) channel was used for acquisition. EOG channel was not used for independent component analysis (ICA) artefact removal, since all the participants were instructed to keep their eyes closed throughout the experiment, except when they were under rest. The subjects were seated on a wooden chair, in a well ventilated room. The subjects were also instructed to rest their feet on a wooden support to ensure that there was no electrical contact with the floor.  
\subsection{Subjects for the study}
A total of seven healthy subjects participated in the experiments (5 males and 2 females, all right handed, mean age of 22.5 years, $\sigma=3.4$). Some of the participants took part in more than one protocol. The protocol was approved by the institute human ethics committee of Indian Institute of Science, Bangalore. The participants signed an informed consent form before taking part in the experiments. Subjects were labelled 1 through 7, and protocols were labelled A through E. Thus, a label `B4', for example, refers to the subject 4 under protocol B.
\subsection{Web application for controlled timing}
For the purpose of maintaining consistent intervals of time for each of the trials of the experiment and to administer cues with as less human error as possible, a web application was developed. The application was programmed to display event labels - ``LEFT'', ``RIGHT'', ``REST'' and ``START''. To prevent the subjects from predicting the sequence of events, the presenting of one of the labels ``LEFT'' and ``RIGHT'' was randomized, with ``REST''/``START'' necessarily following the event label. The duration for each event was also specifically coded-in, depending on the protocol in effect, to ensure that the trials were consistent in duration throughout the experiment. The web-app was used by one of the experiment coordinators in-charge of delivering  cues (auditory/ somatosensory) as per the protocol in effect, as detailed in Section \ref{secp}. A pause button was also made available to facilitate the subjects to choose the number of successive trials before taking rest. 
\begin{figure}
\centering
\includegraphics[width=9cm]{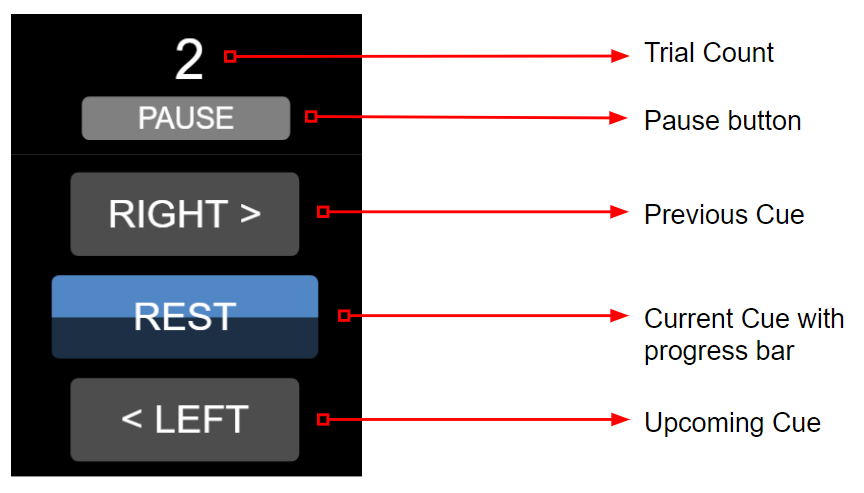}
\caption{A screenshot of the web-app used for timing the auditory cues. The web-app was used to improve the consistency of trial durations.} \label{fig1}
\end{figure}
\subsection{Preprocessing of the EEG data}
The acquired EEG was preprocessed using EEGLAB \cite{delorme2004eeglab}. The 50 Hz line noise was removed using notch filter applied from 49-51 Hz and then, the data was bandpass filtered from 8-30 Hz. This was performed, since the changes in EEG signals due to motor imagery are more visible in mu and beta bands \cite{mcfarland2000mu}\cite{pfurtscheller2006mu}. From the literature, it is known that motor imagery activation in mu and beta bands is visible in the central (Rolandic) region of the brain, primarily consisting of ``C3'',``Cz'' and ``C4'' electrodes. When the arm is moved (or imagined to have moved), mu rhythm changes contralateral to that arm movement (imagination) \cite{ref2}. The frequency band of Rolandic beta rhythms varies from subject to subject, while falling in the broad range of 14-30 Hz \cite{ref3}.

\subsection{Algorithm to identify the primary region of activation}
It is known from the existing literature that for motor imagery, a minimum of 8 channels are needed to obtain optimal performance \cite{ref1}. However, increasing the number of channels introduces the curse of dimensionality. Thus, we use a spatial filtering algorithm that combines several channels into one, using weighted linear combinations, through which features are extracted.

The common spatial pattern (CSP) algorithm linearly transforms the multi-channel EEG signal into a low-dimensional subspace such that the variance of the EEG signal from one class (a group of channels) is maximized while that from the other class (the group of remaining channels) is minimized \cite{jerrinNovel}. Mathematically, the CSP algorithm extremizes the following objective function:
	\begin{equation}
	J(\mathbf{w})=\frac{\mathbf{w}^TX_1X_1^T\mathbf{w}}{\mathbf{w}^TX_2X_2^T\mathbf{w}}=\frac{\mathbf{w}^TC_1\mathbf{w}}{\mathbf{w}^TC_2\mathbf{w}}
	\label{eq1}
	\end{equation}
where $T$ denotes the matrix transpose, $X_i$ is the matrix containing the EEG signals of class $i$, with data samples as columns and channels as rows, $\mathbf{w}$ is the spatial filter and $C_i$ is the spatial covariance matrix of  class $i$.
	
CSP finds spatial filters such that the variance of the filtered data is maximal for one class while simultaneously minimal for the other class. This effectively highlights those regions of the brain that contribute most to a particular action (imagery/actual movement) by comparing it to the activation as obtained during the other action. CSP is especially effective in brain-computer interface (BCI) applications that use oscillatory data like EEG, since their most important features are band-power features. The CSP algorithm is computationally efficient and easy to implement. However, CSP does have some limitations. It is not robust to noise or non-stationarities and may not generalize well to new data, especially when there is very less data available.
Since we are mainly interested in the spatial regions of the brain that are associated with the imagery in consideration, we decided to use the filters produced by the Lagrangian CSP algorithm and implemented it using Fabien Lotte's RCSP-Toolbox for MATLAB \cite{rcsp}. 

\section{Experimental protocols explored}
\label{secp}
The scope of this work was to design a protocol that would return consistent results in terms of the area of activation as expected based on previous work carried out on similar motor imagery tasks. In the literature, all the reported works use either audio or visual cues to the subjects, following which they were expected to imagine the motor action (motor imagery) for the cue given \cite{pfurtscheller2001motor,brandl2015bringing}\cite{brandl2016alternative}. The most popular datasets in use in this domain are the BCI competition datasets \cite{blankertz2004bci}. 
In BCI competition III dataset, the subjects had to imagine either closing the right arm into a fist or wiggling the toes on the right foot. To validate our work, we tested our processing pipeline on the standard datasets available and the obtained results were comparable to those obtained by Lotte and Guan \cite{rcsp}.

The audio and somatosensory cues as detailed in TABLE \ref{table:1} were delivered by one of the experiment conductors seated directly in front of the subject, using the web-app for timing.
After a set of 25-30 trials, the web-app was paused in order to let the subject take rest for approximately 2-3 minutes. Depending on their fatigue level, the subject had a choice to request for less number of trials to be conducted for the next batch of trials.

\subsection{Protocol A: RA-RF with auditory cues}
\begin{figure}
\includegraphics[width=\textwidth]{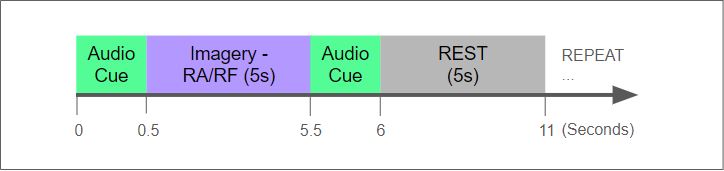}
\caption{Graphical representation of the timings of protocol A. An auditory cue lasting less than 0.5 sec was given to ask the subject to start imagining the motor movement. After a period of 5 sec, another auditory cue was given asking the subject to stop the imagination.}
\label{figa}
\end{figure}

To begin with, the first three subjects (A1,A2,A3) were made to follow the protocol as described in BCI competition III dataset IVa's description as closely as possible. This was to serve as a baseline for the work covered in this paper. The protocol included three types of events - imagery task of right arm (RA) or right foot (RF), and a rest event. Subjects were given rest for 5 seconds after every imagery task, which itself lasted for 5 seconds. The order of the imagery task was randomized to ensure that the subjects were unable to predict the sequence of cues. The graphical representation of the protocol timings is shown in Fig. \ref{figa}. The feedback obtained from the subjects on this protocol was that it was often difficult to imagine the motor action without actually executing it and even harder to focus on the imagery for a duration as long as 5 seconds. The subjects described their imagination of the action merely as a visual image of the action, which does not strictly correspond to the motor imagery/planning that we were looking for. Taking this feedback into account, changes were made to the protocol, giving rise to newer protocols explained below.
\FloatBarrier

\subsection{Protocol B: RA-RF with actual motion and auditory cues}
\begin{figure}
\includegraphics[width=\textwidth]{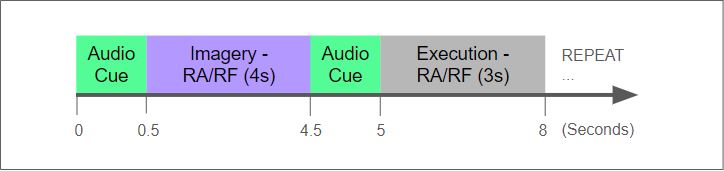}
\caption{Graphical representation of the timings of protocol B. An auditory cue lasting less than 0.5 sec was given to ask the subject to start imagining the motor movement. After a period of 5 sec, another auditory cue was given asking the subject to actually perform the action once.}
\label{figb}
\end{figure}

EEG data of two subjects, B4 and B5 was recorded under this protocol. This protocol has 4 types of events - two imagery tasks of right arm (RA) and right foot (RF), two actual motion tasks of right arm and right foot. Subjects were given auditory cues ``arm" and ``foot" to perform the corresponding imagery tasks repeatedly for 4 sec. Followed by each imagery task, an actual action of the previously imagined task was asked to be performed before 3 sec. The auditory cue for the actual action was ``start". The graphical representation of the protocol timings is shown in Fig. \ref{figb}.

The changes introduced in this protocol were in consideration of the feedback obtained from subjects who participated in protocol A. As seen in section \ref{resprotb}, the introduction of actual movement after the imagery task does produce better results.
\FloatBarrier

\subsection{Protocol C: LA-RA with actual motion and auditory cues}
\begin{figure}
\includegraphics[width=\textwidth]{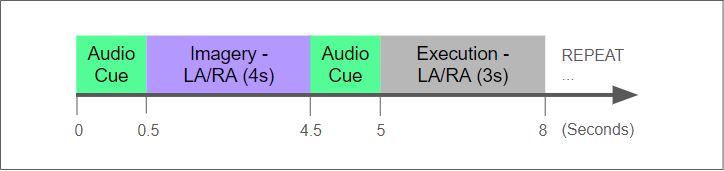}
\caption{Graphical representation of the timings of protocol C. An auditory cue lasting less than 0.5 sec was given asking the subject to start imagining the motor movement. After a period of 5 sec, another auditory cue was given asking the subject to actually perform the action once.}
\label{figc}
\end{figure}

Using this protocol, five subjects' (C1,C3,C4,C5,C6) EEG data were recorded.
This protocol has 4 types of events - two imagery tasks of left arm (LA) and right arm (RA), two actual motion tasks of left arm and right arm. Subjects were given auditory cues ``left" and ``right" to perform the corresponding imagery tasks which lasted for 4s. Followed by each imagery task, an actual action of the previously imagined task was asked to be performed once within 3s. The auditory cue for the actual action was ``start". Since we know that the sensory cortex coincides with the motor cortex in the central region of the brain, we thought introducing localized somatosensory cues will further improve our results and realized the next protocol. The graphical representation of the protocol timings is shown in Fig. \ref{figc}.
\FloatBarrier

\subsection{Protocol D: LA-RA with somatosensory cues}
\begin{figure}
\includegraphics[width=\textwidth]{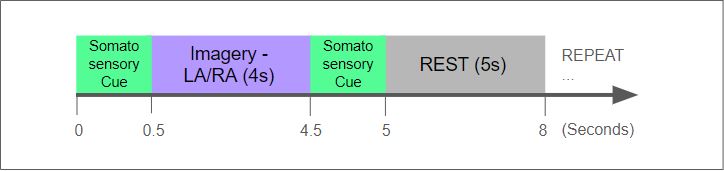}
\caption{Graphical representation of the timings of protocol D. Tap on the wrist was the somatosensory cue for starting the imagination and tap on the knee was the cue for stopping the imagination and taking rest.} \label{figd}
\end{figure}

One subject (D2) was recorded using this protocol.
This protocol has 3 types of events - two imagery tasks of left arm (LA) and right arm (RA) and one rest event. Somatosensory cues were given on the subject's outer wrist in the form of a gentle tap to let the subject know that he/she should perform the imagery task of the corresponding hand. The imagination was asked to be carried out for 4s and a tap on the knee of the same side of the body as the previous cue was given to let the participant stop imagination and take rest. The graphical representation of the protocol timings is shown in Fig. \ref{figd}.  
This protocol was to test if the introduction of somatosensory cues would affect the quality of results obtained. As in section \ref{resprotc}, there is benefit to using somatosensory cues.
\FloatBarrier

\subsection{Protocol E: LA-RA with actual motion and somatosensory cues}
\begin{figure}
\includegraphics[width=\textwidth]{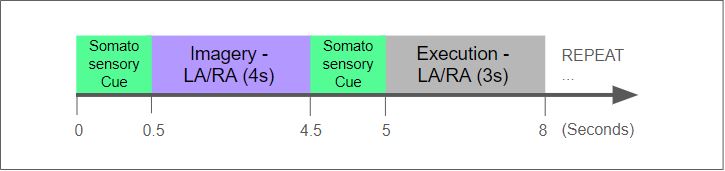}
\caption{Graphical representation of the timings of protocol E. Pressing the inner wrist by the blunt side of a pen was the somatosensory cue for starting the imagination and that of the palm was the cue for performing the actual action.}
\label{fige}
\end{figure}

EEG data of two subjects (E1,E6) were recorded using this protocol. 
This protocol has 4 event types - two imagery tasks of left arm (LA) and right arm (RA), two actual motion tasks of left arm and right arm. The cues that were administered to indicate the onset of imagination or task of actual motion were somatosensory in nature and given using the blunt side of a pen. Upon inducing somatosensory stimulus on the subject's inner wrist, the subject was asked to perform the corresponding imagery task, which lasted for 4s. The somatosensory cue given on the palm told the subject to perform the actual action. The graphical representation of the protocol timings is shown in Fig. \ref{fige}.
\FloatBarrier
\begin{table}[h!]
\caption{List of cues and the corresponding actions.}
\resizebox{\textwidth}{!}{%
\begin{tabular}{|c|c|l|c|}
\hline
\rowcolor[HTML]{C0C0C0} 
Type                                & Cue                                                                                            & Task                                                                                                 & Protocol(s) \\ \hline
Auditory                            & ARM                                                                                            & Imagine closing your right arm into a fist                                                           & A,B         \\ \hline
Auditory                            & FOOT                                                                                           & Imagine wiggling the toes on your right foot                                                         & A,B         \\ \hline
Auditory                            & REST                                                                                           & Take rest                                                                                            & A,B         \\ \hline
Auditory                            & LEFT                                                                                           & Imagine closing your left arm into a fist                                                            & C           \\ \hline
Auditory                            & RIGHT                                                                                          & Imagine closing your right arm into a fist                                                           & C           \\ \hline
Auditory                            & START                                                                                          & \begin{tabular}[c]{@{}l@{}}Perform actual action of previously \\ imagined motor action\end{tabular} & B,C         \\ \hline
Somatosensory                       & Tap on the LEFT inner wrist                                                                    & Imagine closing your left arm into a fist                                                            & D           \\ \hline
Somatosensory                       & Tap on the RIGHT inner wrist                                                                   & Imagine closing your right arm into a fist                                                           & D           \\ \hline
\multicolumn{1}{|l|}{Somatosensory} & Tap on the LEFT/RIGHT knee                                                                     & Take rest                                                                                            & D           \\ \hline
\multicolumn{1}{|l|}{Somatosensory} & \begin{tabular}[c]{@{}c@{}}Poke LEFT inner wrist \\ with the blunt side of a pen\end{tabular}  & Imagine closing your left arm into a fist                                                            & E           \\ \hline
\multicolumn{1}{|l|}{Somatosensory} & \begin{tabular}[c]{@{}c@{}}Poke RIGHT inner wrist \\ with the blunt side of a pen\end{tabular} & Imagine closing your right arm into a fist                                                           & E           \\ \hline
Somatosensory                       & \begin{tabular}[c]{@{}c@{}}Poke LEFT palm with \\ the blunt side of a pen\end{tabular}         & Close your left arm into a fist (actual action)                                                      & E           \\ \hline
Somatosensory                       & \begin{tabular}[c]{@{}c@{}}Poke RIGHT palm with \\ the blunt side of a pen\end{tabular}        & Close your right arm into a fist (actual action)                                                     & E           \\ \hline
\end{tabular}%
}
\label{table:1}
\end{table}

\section{Results}
We compare the performance of the tested protocols by visually inspecting the filters produced by the CSP algorithm. The criteria of evaluation is the correctness of regions highlighted in the filters, with reference to what we know about the expected activation of the brain for the motor imagery in consideration. It may be noted that the sign of the elements in the spatial filer vectors are not significant and only the absolute value matters. Also, all the topoplots were obtained using the EEG data corresponding to motor imagery alone, although some protocols had actual movements included in them. There are broadly two types of imagery in the protocols tested - arm and foot.
\subsection{RA-RF based protocols}
 According to literature \cite{rcsp}, the CSP filter plots for clenching of right arm into a fist when compared to wiggling of right toes on the foot (RA-RF protocol) indicate activation on the left hemisphere of the brain along electrodes ``C3'', ``C5'' and ``CP3''. The CSP filter plots for wiggling of right toes on the foot when compared to clenching of right arm is along the central electrode ``Cz''.

\begin{figure}
  \centering
    \includegraphics[width=9cm]{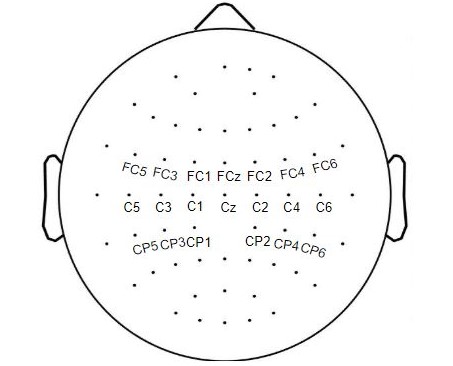}
    \caption{10/10 EEG electrode placement system. The dots correspond to the positions of the electrodes on the scalp. The names of the relevant electrodes are also given.}
\end{figure}

\subsubsection{Protocol A: RA-RF with auditory cues}
\begin{figure}[h!]
  \centering
  \subfloat[]{
    \includegraphics[width=6cm]{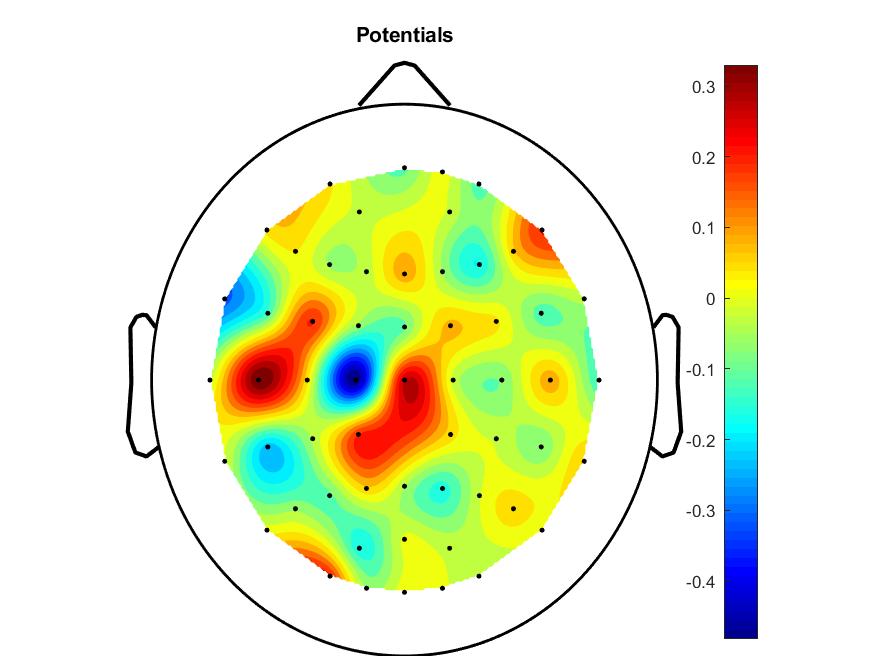}}
  \subfloat[]{
    \includegraphics[width=6cm]{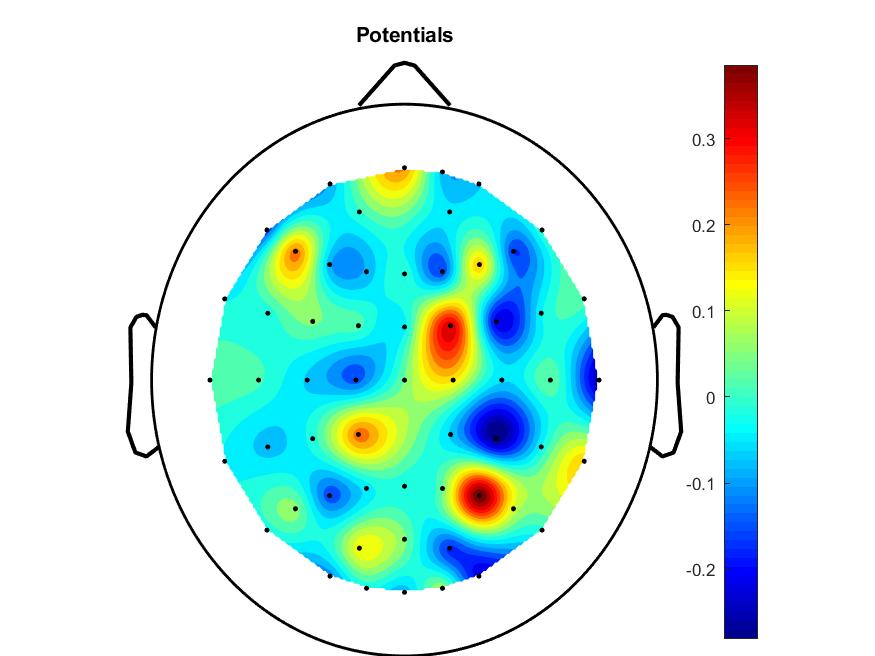}}
  \caption{CSP filter obtained for subject A3 for (a) right arm and (b) right foot motor imageries under protocol A. Under this protocol, three events were there, namely, imagery of right arm movement, imagery of right foot movement and rest state.}
     \label{fig8}
\end{figure}

As shown in Figs. \ref{fig8} (a) and (b), the topoplots obtained  highlight the regions as expected for the imagery pair RA-RF. However, the regions are not defined very distinctly. This was an observation common to all subjects who participated in this protocol. This could be because the subjects found it difficult to imagine the motor action without actually performing the action. This is also likely affected by the fact that the subjects were unable to stay focused on the imagery for extended periods of time. 
Taking these into account, changes were made to the protocol.

\subsubsection{Protocol B: RA-RF with auditory cues, followed by actual motion}\label{resprotb}
\begin{figure}[h!]
  \centering
  \subfloat[]{
    \includegraphics[width=6cm]{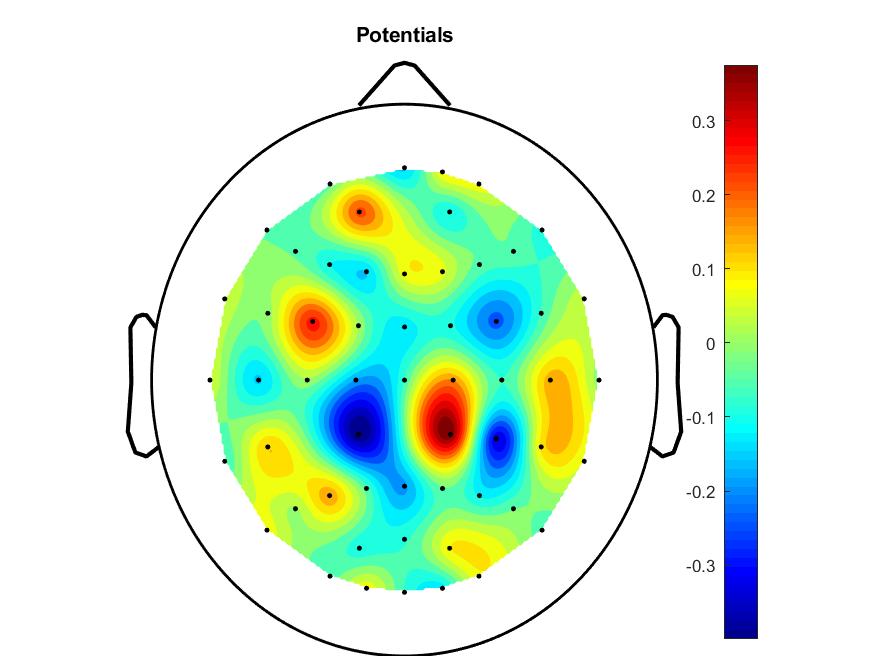}}
  \subfloat[]{
    \includegraphics[width=6cm]{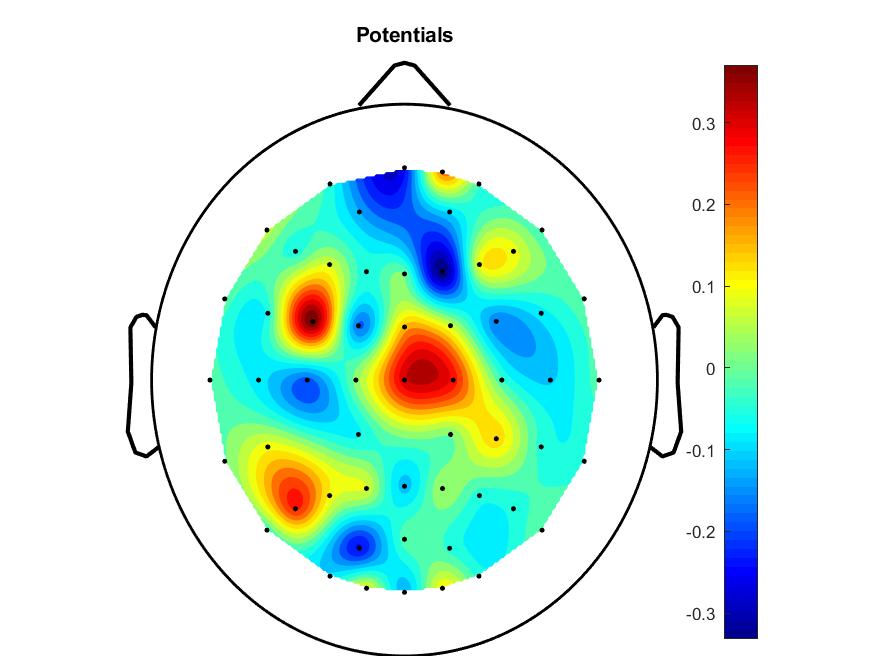}}
  \caption{CSP filter obtained for subject B4 for (a) right arm and (b) right foot motor imageries, under protocol B. Under this protocol, four events were there, namely, imagery of right arm and right foot movement, and actual movement of right arm and right foot.}
  \label{fig9}
\end{figure}

As seen in Fig. \ref{fig9}, the introduction of actual motion into the protocol following every single trial of imagery did result in better results. The topoplots exhibit more definitive regions and the subjects also reported that it was easier to stay focused through the trials, since they were instructed to perform the actual movement.

\subsection{LA-RA based protocols}
For the clenching of a hand into a fist, we expect activation predominantly centered around the central line electrodes (``C5'',``C3'', ``C1'', ``Cz'', ``C2'', ``C4'' and ``C6''). This activation is also known to show contralateralization with respect to the left or right arm in imagery. In other words, for imagery related to the left arm, the right side of the brain is activated more and vice-versa for the right hand. This property of lateral symmetry of activity for imagery of the left or right arm helps in evaluating visually, the performance of the protocols in question.

\subsubsection{Protocol C: LA-RA with auditory cues, followed by actual motion}\label{resprotc}
\begin{figure}[h!]
  \centering
  \subfloat[]{
    \includegraphics[width=6cm]{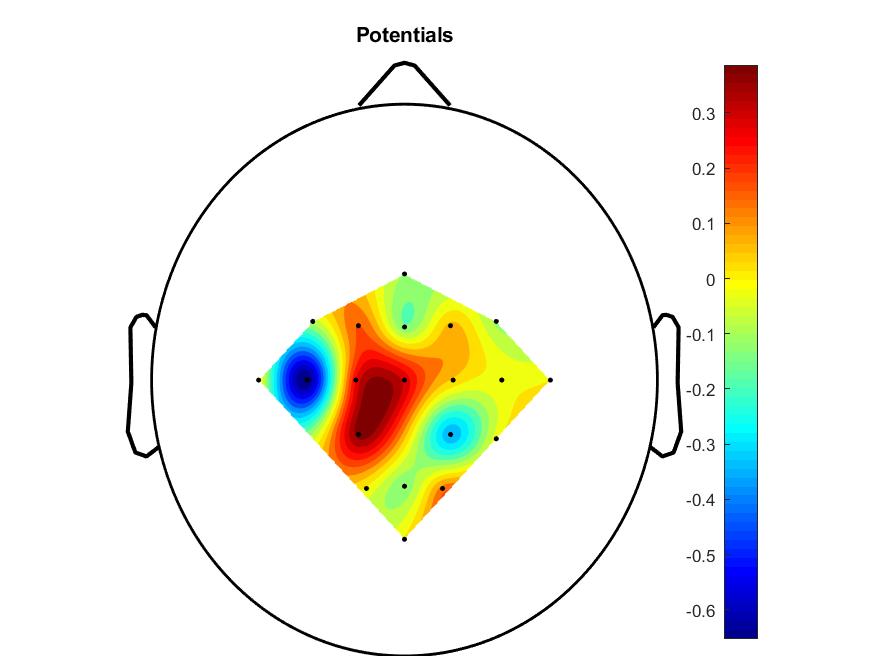}}
  \subfloat[]{
    \includegraphics[width=6cm]{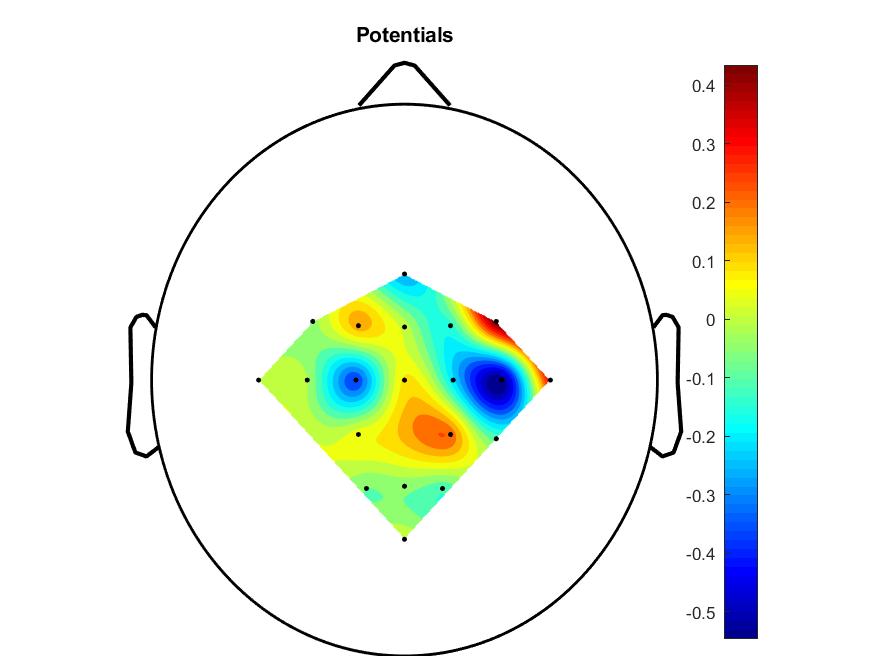}}
      \caption{CSP filter obtained for subject C6 for (a) right arm and (b) left arm motor imageries, under protocol C. Under this protocol, four events were there, namely, imagery of right arm and right foot movement, and actual movement of right arm and right foot. Also,  only a reduced set of electrodes were considered for this protocol.}
      \label{fig10}
\end{figure}

For this protocol and those to follow, a reduced electrode set was considered to reduce dimensions. This was decided based on our understanding of the regions within which the two motor imagery (left arm and right arm) are expected to lie. The topoplots of the filters (Fig. \ref{fig10}), obtained with this protocol, are very close to the ones reported in the literature.
\FloatBarrier
\subsubsection{Protocol D: LA-RA with somatosensory cues}
\begin{figure}[h!]
  \centering
    \subfloat[]{
    \includegraphics[width=6cm]{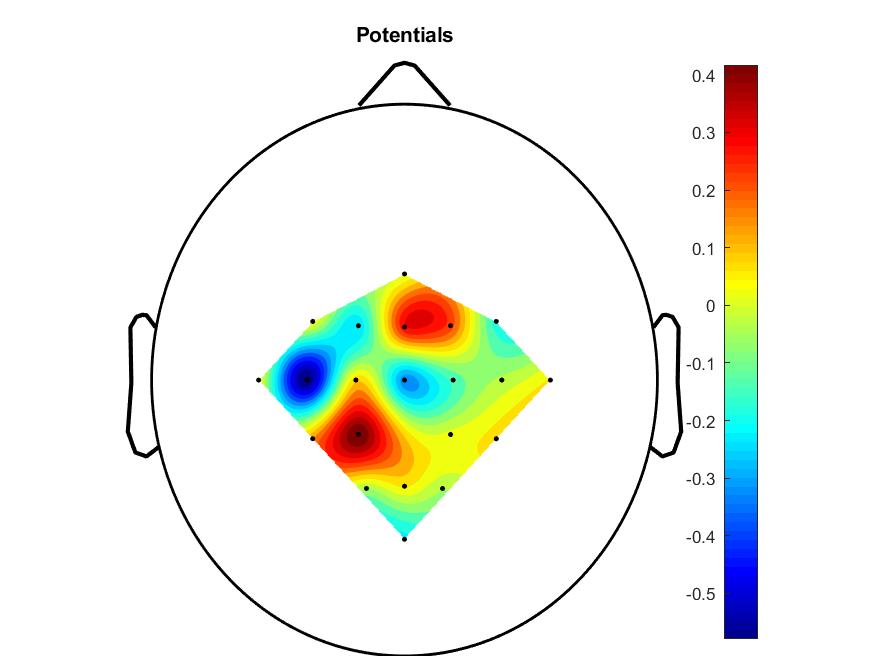}}
   \subfloat[]{
    \includegraphics[width=6cm]{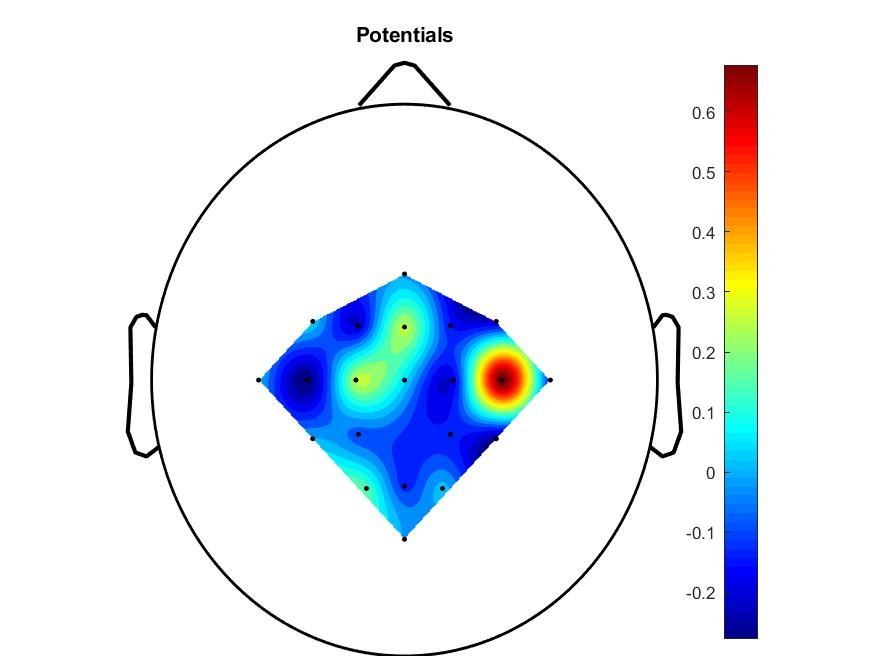}}
      \caption{CSP filter obtained for subject D2 for (a) right arm and (b) left arm motor imageries, under protocol D. Under this protocol, three events were there, namely, imagery of right arm and right foot movement, and rest state. Instead of auditory cues, localized somatosensory cues were used.   Also,  only a reduced set of electrodes were considered for this protocol.}
          \label{fig11}
\end{figure}

As seen in the topoplots in Fig. \ref{fig11}, the introduction of localized somatosensory cues as detailed earlier helps the subject to stay focused during the imagery and leads to better results.
\FloatBarrier
\subsubsection{Protocol E: LA-RA with somatosensory cues, followed by actual motion}
\begin{figure}[h!]
  \centering
  \subfloat[]{
    \includegraphics[width=6cm]{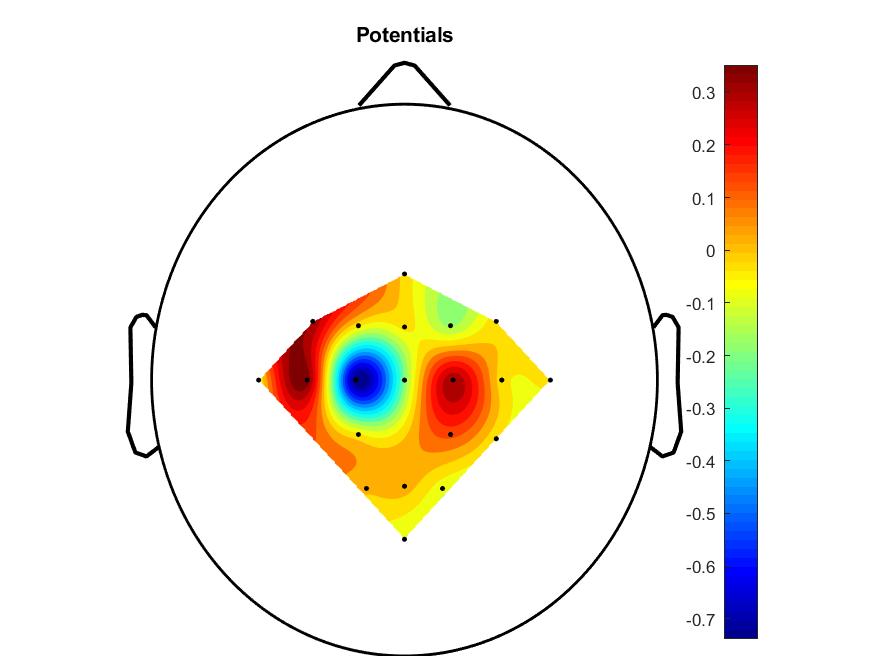}}
  \subfloat[]{
    \includegraphics[width=6cm]{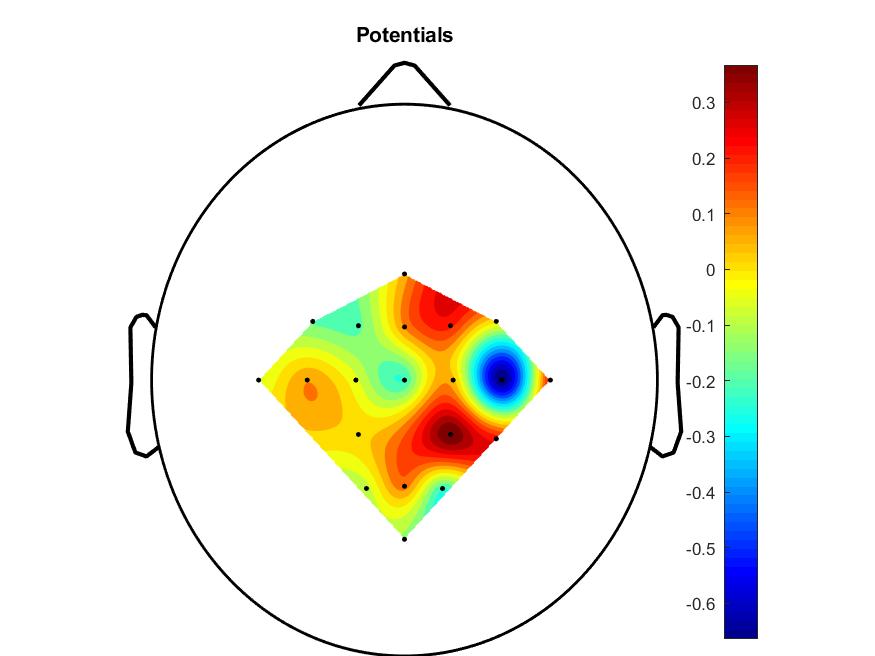}}
      \caption{CSP filter obtained for subject E6 for (a) right arm  and (b) left arm motor imageries, under protocol E. Under this protocol, four events were there, namely, imagery of right arm and right foot movement, and actual movement of right arm and right foot. Instead of auditory cues, localized somatosensory cues were used in this protocol. Also,  only a reduced set of electrodes were considered for this protocol.}
          \label{fig12}
\end{figure}

Considering the benefits of somatosensory cues from protocol D, protocol E was brought about to put protocols C and D together. This was tested on 2 of the 5 subjects, who participated in protocol C. The improvements do seem to stack up as seen in the topoplots for this protocol (Fig. \ref{fig12}). The activation is exactly where they are expected and the plots themselves have sharp and defined regions, which are considered favorable traits. 
\FloatBarrier

\section{Discussion}
With four different segments of protocols, we have found CSPs for motor imagery with auditory cues, with auditory cues followed by actual movement, somatosensory cues and, somatosensory cues followed by actual movement. It is interesting to note that with somatosensory cues, the regions getting activated for motor imagery are more focused as is evident from the CSP topoplots. The regions activated are in sync with the previous experiments based on EEG \cite{pfurtscheller2001motor,mcfarland2000mu} and fMRI \cite{hanakawa2003functional,sharma2008mapping} to localize motor imagery activity. 

In our experiment, motor imagery with auditory cues for right foot was concentrated more closely in the center in topoplot (near ``Cz''). This result is in sync with functional near infrared spectroscopy (fNIR) study \cite{batula2017comparison} where bilateral activations were visible for limb movement imagery. Saimpont et al. have reported that auditory cues are helpful for greater visualization of motor imagery \cite{saimpont2013motor}. The regions activated in our study are intraparietal sulcus (IPS), supramarginal gyrus and precentral sulcus which were also activated during motor imagery in fMRI based experiment \cite{sharma2008mapping}. The increased activity of medial prefrontal cortex (mPFC) (Sensorimotor area) has been reported for motor imagery. Significant ipsilateral activations were seen in brain region represented by ``P4''/``CP4'' channels in recovery patients of motor neuron disease \cite{bundy2017contralesional}. Somatosensory cues were shown to improve motor imagery performance measured with motor evoked potentials \cite{bonassi2017provision}.

Anterior and posterior Brodmann area (BA) 4 are involved during motor imagery as reported earlier in fMRI study on healthy subjects \cite{sharma2008mapping}. However, with 64 channel EEG aquisition systems, it is hard to report focused cluster-based activation in BA 4 due to the low spatial resolution.  
\section{Conclusions}
The results obtained certainly show promise. Changing the protocol to include actual movement following the imagery helps improve the results. This arises from the fact that healthy subjects are unlikely to have been in a situation where their brain exercised motor planning, (i.e, attempted to move a certain limb) but was unable, or did not want to move the limb. Thus, healthy subjects find it difficult to imagine only the planning, without actually performing the associated movement later. On the other hand, allowing the subjects to move after the trial ensures that the trial captures the activity of the brain during planning. 

The results also seem to indicate that somatosensory cues stimulate the subjects better, which results in more focused motor imagery. With evolutionary perspective, somatosensory and motor areas are in close association, which leads to diffused activation in pre and post central gyrus areas. This helps not only in inducing activity in the same part of the brain, but also makes it easier for the subject to localize. A subject, who is touched on his hand, would be able to easily resolve where he was touched and will be able to quickly imagine motor actions on the same hand. This resolution based on localization, we speculate, results in better quality of motor imagery.
\section{Limitations}

Due to approximation of locations in EEG cap, signals from ``C3'' and ``C4'' reflect the activity from both the motor and the somatosensory cortex. Making a distinction between motor and somatosensory activity was not feasible with EEG. 

%
%
\bibliographystyle{splncs04}
\bibliography{name}

\begin{thebibliography}{10}
\providecommand{\url}[1]{\texttt{#1}}
\providecommand{\urlprefix}{URL }
\providecommand{\doi}[1]{https://doi.org/#1}

\bibitem{batula2017comparison}
Batula, A.M., Mark, J.A., Kim, Y.E., Ayaz, H.: Comparison of brain activation
  during motor imagery and motor movement using fnirs. Computational
  intelligence and neuroscience  \textbf{2017} (2017)

\bibitem{blankertz2004bci}
Blankertz, B., M{\"u}ller, K.R., Curio, G., Vaughan, T.M., Schalk, G., Wolpaw,
  J.R., Schl{\"o}gl, A., Neuper, C., Pfurtscheller, G., Hinterberger, T.,
  et~al.: The bci competition 2003. IEEE Trans. Biomed. Eng  \textbf{51}(6),
  1044--51 (2004)

\bibitem{bonassi2017provision}
Bonassi, G., Biggio, M., Bisio, A., Ruggeri, P., Bove, M., Avanzino, L.:
  Provision of somatosensory inputs during motor imagery enhances
  learning-induced plasticity in human motor cortex. Scientific reports
  \textbf{7}(1), ~9300 (2017)

\bibitem{brandl2015bringing}
Brandl, S., H{\"o}hne, J., M{\"u}ller, K.R., Samek, W.: Bringing bci into
  everyday life: Motor imagery in a pseudo realistic environment. In: 2015 7th
  International IEEE/EMBS Conference on Neural Engineering (NER). pp. 224--227.
  IEEE (2015)

\bibitem{brandl2016alternative}
Brandl, S., M{\"u}ller, K.R., Samek, W.: Alternative csp approaches for
  multimodal distributed bci data. In: 2016 IEEE International Conference on
  Systems, Man, and Cybernetics (SMC). pp. 003742--003747. IEEE (2016)

\bibitem{bundy2017contralesional}
Bundy, D.T., Souders, L., Baranyai, K., Leonard, L., Schalk, G., Coker, R.,
  Moran, D.W., Huskey, T., Leuthardt, E.C.: Contralesional brain--computer
  interface control of a powered exoskeleton for motor recovery in chronic
  stroke survivors. Stroke  \textbf{48}(7),  1908--1915 (2017)

\bibitem{claassen2019detection}
Claassen, J., Doyle, K., Matory, A., Couch, C., Burger, K.M., Velazquez, A.,
  Okonkwo, J.U., King, J.R., Park, S., Agarwal, S., et~al.: Detection of brain
  activation in unresponsive patients with acute brain injury. New England
  Journal of Medicine  \textbf{380}(26),  2497--2505 (2019)

\bibitem{delorme2004eeglab}
Delorme, A., Makeig, S.: {EEGLAB}: an open source toolbox for analysis of
  single-trial {EEG} dynamics including independent component analysis. Journal
  of neuroscience methods  \textbf{134}(1),  9--21 (2004)

\bibitem{hanakawa2003functional}
Hanakawa, T., Immisch, I., Toma, K., Dimyan, M.A., Van~Gelderen, P., Hallett,
  M.: Functional properties of brain areas associated with motor execution and
  imagery. Journal of neurophysiology  \textbf{89}(2),  989--1002 (2003)

\bibitem{jerrinNovel}
Jerrin, T.P., Ramakrishnan, A., Ananthapadmanabha, T.: A novel deep learning
  architecture for decoding imagined speech from {EEG}. In: IEEE Austria
  International Biomedical Engineering Conference (AIBEC 2019). IEEE (2019)

\bibitem{ref3}
Kropotov, J.D.: Quantitative {EEG}, event-related potentials and neurotherapy.
  Academic Press (2010)

\bibitem{ref2}
Kropotov, J.D.: Functional neuromarkers for psychiatry: Applications for
  diagnosis and treatment. Academic Press (2016)

\bibitem{rcsp}
Lotte, F., Guan, C.: Regularizing common spatial patterns to improve {BCI}
  designs: unified theory and new algorithms. IEEE Transactions on biomedical
  Engineering  \textbf{58}(2),  355--362 (2010)

\bibitem{mcfarland2000mu}
McFarland, D.J., Miner, L.A., Vaughan, T.M., Wolpaw, J.R.: Mu and beta rhythm
  topographies during motor imagery and actual movements. Brain topography
  \textbf{12}(3),  177--186 (2000)

\bibitem{pfurtscheller2006mu}
Pfurtscheller, G., Brunner, C., Schl{\"o}gl, A., Da~Silva, F.L.: Mu rhythm (de)
  synchronization and {EEG} single-trial classification of different motor
  imagery tasks. NeuroImage  \textbf{31}(1),  153--159 (2006)

\bibitem{pfurtscheller2001motor}
Pfurtscheller, G., Neuper, C.: Motor imagery and direct brain-computer
  communication. Proceedings of the IEEE  \textbf{89}(7),  1123--1134 (2001)

\bibitem{saimpont2013motor}
Saimpont, A., Malouin, F., Tousignant, B., Jackson, P.L.: Motor imagery and
  aging. Journal of motor behavior  \textbf{45}(1),  21--28 (2013)

\bibitem{ref1}
Sannelli, C., Dickhaus, T., Halder, S., Hammer, E.M., M{\"u}ller, K.R.,
  Blankertz, B.: On optimal channel configurations for {SMR}-based
  brain--computer interfaces. Brain topography  \textbf{23}(2),  186--193
  (2010)

\bibitem{sharma2008mapping}
Sharma, N., Jones, P.S., Carpenter, T., Baron, J.C.: Mapping the involvement of
  {BA} 4a and 4p during motor imagery. Neuroimage  \textbf{41}(1),  92--99
  (2008)

\end{thebibliography}
\end{document}